\documentclass[conference]{IEEEtran}
\usepackage{cite,booktabs,graphicx,psfrag,subfigure,mathtools}
\usepackage{url,amsmath,array,amssymb,amsfonts,hyperref,cite,epstopdf}
\input{epsf.sty}
\usepackage{epsfig}

\newtheorem{definition}{Definition}

\newtheorem{example}{Example}

\title{Decoding Network Codes using the Sum-Product Algorithm}
\begin{document}

\author{
\authorblockN{Anindya Gupta and B. Sundar Rajan}
\authorblockA{Dept. of ECE, IISc, Bangalore 560012, India, Email:\{anindya.g, bsrajan\}@ece.iisc.ernet.in}
}

\maketitle
\thispagestyle{empty}	
\begin{abstract}
While feasibility and obtaining a solution of a given network coding problem are well studied, the decoding procedure and complexity have not garnered much attention. We consider the decoding problem in a network wherein the sources generate multiple messages and the sink nodes demand some or all of the source messages. We consider both linear and non-linear network codes over a finite field and propose to use the sum-product (SP) algorithm over Boolean semiring for decoding at the sink nodes in order to reduce the computational complexity. We use \textit{traceback} to further lower the computational cost incurred by SP decoding. We also define and identify a sufficient condition for \textit{fast decodability} of a network code at a sink that demands all the source messages.
\end{abstract}
\section{Introduction}
In contemporary communication networks, the nodes perform only routing, \textit{i.e.}, they copy the data on incoming links to the outgoing links. In order to transmit messages generated simultaneously from multiple sources to multiple sinks the network may need to be used multiple times. This limits the throughput of the network and increases the time delay too. It is known that if intermediate nodes in a network are permitted to perform coding operations, \textit{i.e.}, encode data received on the incoming links and then transmit it on the outgoing links (each outgoing link can get differently encoded data), the throughput of the network increases. This is called network coding \cite{Yeung}. Thus, network coding subsumes routing.

\begin{figure}[h]
\centering
\includegraphics[scale=0.5]{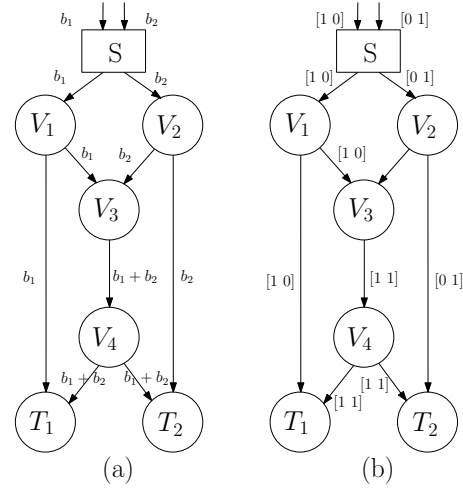}
\caption{Butterfly Network (a) A network code and (b) Global encoding vectors}
\label{b_fly}
\vspace{-14pt}
\end{figure}

For example, consider the butterfly network \cite{Yeung} of Fig. \ref{b_fly} wherein each link can carry one bit per link use, source node $S$ generates bits $b_1$ and $b_2$, and both sink nodes $T_1$ and $T_2$ demand both source bits. With routing only, two uses of link $V_3-V_4$ are required while with network coding only one. 

Above is an example of single-source multi-sink linear multicast network coding, wherein there is a single source ($S$), generating a finite number of messages, ($x_1,x_2$), and multiple sinks, each demanding all the source messages and the encoding operations at all nodes are linear. In general, there may be several source nodes, each generating different number of source messages, and several sink nodes, each demanding only a subset, and not all, of source messages. Decoding at sink nodes with such general demands is studied in this paper.

We represent a network by a finite directed acyclic graph $N=(V,E)$, where $V$ is the set of vertices or nodes and $E \subseteq V \times V$ is the set of directed links or edges between nodes. All links are assumed to be error-free. Let $[n]=\{1,2,\ldots,n\}$. The network has $J$ sources, $S_j,\,j\in [J]$, and $K$ sinks, $T_k,\,k\in [K]$. The source $S_j$ generates $\omega _j$ messages, $\forall j\in [J]$. Let $\omega=\sum _{j=1}^J \omega _j$ be the total number of source messages. The $\omega$-tuple of source messages is denoted by $x_{[\omega]}=(x_1,x_2,\ldots,x_\omega)$, where $x_i\in F,\,\forall i\in [\omega]$ and $F$ is a finite field. By $\mathbf{x}=(x_1,\ldots,x_{\omega})^T$ we denote the column vector of the source messages. The demand of the $k^{th}$ sink node is denoted by $D_k\subseteq [\omega]$. Given a set $I=\{i_1\ldots,i_l\}\subseteq [\omega]$, let $x_I=(x_{i_1},\ldots,x_{i_l})$, \emph{i.e.}, $x_{[\omega]}$ \emph{restricted} to $I$. Let $\{x_I\}=\{x_I\colon x_I\in F^I\}$, \emph{i.e.}, the set of all $I$-tuples over $F$. For a multi-variable binary-valued function $f(x_1,\ldots,x_\omega)$, the subset of $F^\omega$ whose elements are mapped to $1$ by $f(x_1,\ldots,x_\omega)$ is called its support and is denoted by  $\mathtt{supt}(f(x_{[\omega]}))$ and $\mathtt{supt}_{I}(f(x_{[\omega]}))$ denotes the $|I|$-tuples in the support restricted to $I$. A source message is denoted by edges without any originating node and terminating at a source node. Data on a link $e\in E$ is denoted by $y_e$. 

A network code (NC) is a set of coding operations to be performed at each node such that the requisite source messages can be faithfully reproduced at the sink nodes. It can be specified using either local or global description \cite{Yeung}. The former specifies the data on a particular outgoing edge as a function of data on the incoming edges while the latter specifies the data on a particular outgoing edge as a function of source messages. Throughout the paper we use global description for our purposes. 
\begin{definition}
Global Description of an NC \cite{Yeung}: An $\omega$-dimensional NC on an acyclic network over a field $F$ consist of $|E|$ global encoding maps $\tilde{f}_e : F^{\omega}\rightarrow F, \forall e\in E$ (i.e.,  $\tilde{f}_e ( \mathbf{x}) = y_e$).
\end{definition}
Let $e_i,i=1,\ldots,\omega$, be the incoming edges at the source, then $y_{e_i}=x_i$. 

When the intermediate nodes perform only linear encoding operations then such an NC is said to be a linear network code (LNC).
\begin{definition}
Global Description of an LNC \cite{Yeung}: An $\omega$-dimensional LNC on an acyclic network over a field $F$ consist of $|E|$ $1\times\omega$ global encoding vectors $\mathbf{f}_e, ~ \forall e\in E$ such that $\mathbf{f}_e\cdot \mathbf{x}=y_e$.
\end{definition}
The global encoding vectors for the incoming edges at the source are standard basis vectors for the vector space $F^{\omega}$. The global encoding vectors of the LNC for butterfly network is given in Fig.~\ref{b_fly}(b). 

Hereafter we assume that the network is feasible, \emph{i.e.}, demands of all sink nodes can be met using network coding and the global description of a network code (linear or non-linear) is given. If a sink node demands $\omega'$ ($\leq$ $\omega$) source messages, it will have at least $\omega'$ incoming edges. The decoding problem is to reproduce the desired source messages from the coded data received at the incoming edges. Thus, decoding amounts to solving for a specified set of $\omega'$ unknowns using a set of at least $\omega'$ simultaneous equations in $\omega$ unknowns. Hence, the global description of the NC is more useful for decoding. 

While decoding of non-linear NC has not been studied, the common technique used for decoding a LNC is to perform Gaussian elimination \cite{Lun1,Lun2}, which requires $\mathcal{O}(n^3)$ operations, followed by backward substitution, which requires $\mathcal{O}(n^2)$ operations ($n$ is the number of variables) \cite{Num}. This is not recommendable when the number of equations and/or variables is very large. In such cases, iterative methods are used. Convergence and initial guess are some issues that arise while using iterative methods \cite{Strang}. 

We propose to use the sum-product (SP) algorithm to perform iterative decoding at the sinks. A similar scheme for decoding multicast network codes using factor graph \cite{Frey} was studied in \cite{Salmond}. The authors considered the case of LNC. The problems associated with the proposed decoding scheme in \cite{Salmond} are:
\begin{itemize}
\item In order to construct the factor graph, full knowledge of network topology is assumed at the sinks which is impractical if the network topology changes. For a particular sink node (say $T$), the factor graph constructed will have $\omega +|E|$ variable nodes and $|E|+ |In(T)|$ factor nodes, where $In(T)$ is the set of incoming edges at node $T$.
\item Complete knowledge of local encoding matrix \cite{Yeung} of each node is assumed at the sinks which again is impractical since local encoding matrix for different nodes will have different dimensions and hence variable number of overhead bits will be required to communicate to downstream nodes which will incur huge overhead.
\end{itemize}
We also point out that the motivating examples, \textit{viz.}, Examples 1 and 4, given in \cite{Salmond} for which the proposed decoding method claims to exploit the network topology admits a simple routing solution and no network coding is required to achieve maximum throughput. Solving linear equations in boolean variables is also studied in \cite{Mez}.

The contributions and organization of the paper are as follows:
\begin{itemize}
\item In Section III-A we pose the problem of decoding of linear and non-linear NC as \emph{marginalize a product function problem} (MPF) and construct factor graph using the global description of network codes. For a particular sink node, the constructed graph will have fewer vertices than in \cite{Salmond} and hence the number of messages and operations performed will also be fewer. Unlike in \cite{Salmond}, our scheme requires only the knowledge of global encoding maps/vectors of incoming edges at a sink node and not the entire network structure and coding operation performed at each node. 
\item In Sections III-B, we utilize \textit{traceback} instead of running multiple-vertex version of algorithm, thus, further reducing the number of operations. Some examples illustrating the proposed techniques are given in Section III-D.
\item We discuss utility and computational complexity of the proposed technique in Section III-C. For sink nodes which demand all the source messages, the notion of \emph{fast decodable network codes} is defined and a sufficient condition for the same is identified. 
\end{itemize}
We present a brief overview of the SP algorithm in Section II and conclude the paper with a discussion on scope for further work in Section IV.

\section{The Sum-Product Algorithm and Factor Graphs}
In this section, we review the computational problem called the MPF problem and specify how SP algorithm can be used to efficiently solve such problems. An equivalent method to efficiently solve MPF problems is given in \cite{Aji} and is called the \emph{generalized distributive law} (GDL) or the \emph{junction tree algorithm}. The simplest example of SP algorithm offering computational advantage is the distributive law on real numbers, $a\cdot (b+c)=a\cdot b+a\cdot c$; the left hand side of the equation requires fewer operation than the right hand side. Generalization of addition and multiplication is what is exploited by the SP (or the junction tree) algorithm in different MPF problems. The mathematical structure in which these operations are defined is known as commutative semirings. 

\begin{definition}
A commutative semiring \cite{Aji} is a set $R$, together with two binary operations ``$+$'' (\textit{addition}) and ``$\cdot$'' (\textit{multiplication}), which satisfy the following axioms:
\begin{enumerate}
\item The operation ``$+$'' satisfies closure, associative, and commutative properties; and there exists an element ``$0$'' (\textit{additive identity}) such that $r+0=r,\forall r\in R$.
\item The operation ``$\cdot$'' satisfies closure, associative, and commutative properties; and there exists an element ``$1$'' (\textit{multiplicative identity}) such that $r\cdot 1=r,\;\forall r\in R$.
\item The operation ``$\cdot$'' \textit{distributes} over ``$+$'', \textit{i.e.}, $r_1\cdot r_2 + r_1\cdot r_3 = r_1\cdot (r_2+r_3),\;\forall r_1,r_2,r_3\in R$
\end{enumerate}
\end{definition}

For different problems, we use different semirings with different notion of ``$+$ and $\cdot$''. Some examples are listed below.
\begin{enumerate}
\item Application of the SP algorithm to Fourier transform yields the FFT algorithm; the semiring is the set of complex numbers with the usual addition and multiplication \cite{Frey,Aji}.
\item ML decoding of binary linear codes is also an MPF problem and application of SP algorithm yields the Gallager-Tanner-Wiberg decoding algorithm over a Tanner graph; the semiring is the set of positive real numbers with ``$\min$'' as sum and ``$+$'' as product, called the min-sum semiring \cite{Frey,Aji}. The BCJR algorithm for decoding turbo codes and LDPC deocoding algorithm are some other applications of SP algorithm.
\item Application to the ML sequence estimation, for instance in decoding convolutional codes, yields the Viterbi algorithm \cite{Aji}; the semiring is again the min-sum semiring.
\item Recently, the GDL has been shown to reduce the ML decoding complexity of space-time block codes in \cite{LP}; the semiring applicable is the min-sum semiring of complex number. The authors introduced \textit{traceback} for GDL and used it to further lower the number of operations. 
\end{enumerate}

Thus, both these algorithms subsume as special cases many well known algorithms. 

\subsection{MPF Problems in Boolean Semiring}
A Boolean semiring is the set $\{0,1\}$ together with the usual Boolean operations $\vee$ (OR) and $\wedge$ (AND). We denote it by $R=(\{0,1\},\vee,\wedge)$. The elements $0$ and $1$ are the \emph{additive} and \emph{multiplicative identities} respectively. The MPF problem defined for this semiring is described below. Let $x_i,\,i\in [n]$ be $n$ variables taking values in finite alphabets $A_i,i\in [n]$. For $I=\{i_1,\ldots,i_k\}\subseteq [n]$, let $x_I=(x_{i_1},\ldots,x_{i_k})$ $A_I=A_{i_1}\times \ldots \times A_{i_k}$. Let $\mathcal{S}=\{S_1,S_2,\ldots,S_M\},\; S_j\subseteq [n]$, such that for each $j\in [M]$, there is a function $\alpha _j:A_{S_j}\rightarrow R$. The functions $\alpha _j$s are called the \emph{local kernels},  the set of variables in $x_{S_j}$ is called the \emph{local domain} associates with $\alpha _j$ and $A_{S_j}$ is the associated \emph{configuration space}. The \emph{global kernel}, $\beta: A_{[n]}\rightarrow R$ and its $i^{th}$ \emph{marginalization}, $\beta _i: A_{S_i}\rightarrow R$, are defined below.
\begin{align} \nonumber
\beta(x_1,x_2,\ldots,x_n)=\bigwedge _{j=1}^M \alpha _j(x_{S_j})  
\end{align}

\begin{align} 
\label{eqn_marg}
\beta_i(x_{i}) = \bigvee _{\{x_{[n]\backslash i}\}} \beta (x_1,x_2,\ldots,x_n)  
\end{align}
\subsection{The SP Algorithm}
Brute force computation of marginalizations \eqref{eqn_marg} require $\mathcal{O}(A_{[n]})$ computations; the SP algorithm is an efficient way of computing these. It involves iteratively passing \emph{messages} along the edges of the \emph{factor graph}, $\mathcal{G}=(\mathcal{V\cup F},\mathcal{E})$, associated with the given MPF problem. The factor graph is a bipartite graph. Vertices in $\mathcal{V}$ are called variable nodes; one for each variable $x_i$, $\forall i\in [n]$ ($|\mathcal{V}|=n$).  The vertices in $\mathcal{F}$ are called the factor nodes; one for each local kernel $\alpha _j$, $\forall j\in [M]$ ($|\mathcal{F}|=M$). A variable node $x_i$ is connected to a factor node $\alpha _j$ iff $x_i$ is an argument of $\alpha _j$. For convenience we assume that for a variable node $x$ the local domain and local kernel are $x$ and $1$ respectively.

Let $N(x_i)$ denote the set of factor nodes adjacent to the variable node $x_i$, \emph{i.e.}, set of local kernels with $x$ as an argument and $N(\alpha _j)(=x_{S_j})$ denote the set of variable nodes adjacent to the factor node $\alpha _j$, \emph{i.e.}, the local domain of $\alpha _j$. The directed message passed from a variable node $x_i$ to an adjacent factor node $\alpha _j$ and vice versa are as follows:
\begin{align}
\label{eqn_msg1}
\mu _{x_i\rightarrow \alpha _j}(x_i)= \bigwedge _{\alpha '\in N(x_i)\backslash \alpha _j}\mu _{\alpha '\rightarrow x_i} (x_i)
\end{align}
\begin{align}
\mu _{\alpha _j \rightarrow x_i}= \bigvee _{\{x_{S_j\backslash i}\}} \alpha _j(x_{S_j}) \bigwedge _{x'\in \{x_{S_j\backslash i}\}}\mu _{x'\rightarrow \alpha _j}(x')
\end{align}

Depending on the requirement, we may need to evaluate marginal(s) at only one, a few or all variable nodes. The algorithm starts at the leaf nodes (nodes with degree one) with the leaves passing messages to the adjacent nodes. Once a vertex has received messages from all but one of its neighbor, it computes its own message and passes it to the neighbor from which it has not yet received the message. This message passing terminates when all the variable nodes at which marginals are required to be evaluated have received from all its neighbors. A node after receiving messages from all of its neighbors, computes its \emph{state} as the product of messages received from all the adjacent nodes. For a variable node, $x_i$, it is denoted by $\sigma_i(x_i)$ and is given as follows:
\begin{align} 
\label{eqn_state_v}
\sigma_i(x_i) = \bigwedge _{\alpha ' \in N(x_i)}\mu_{\alpha ' \rightarrow x_i}(x_i),
\end{align}
Similarly, the state for a factor node is computed as follows:
\begin{align} 
\label{eqn_state_f}
\sigma_{\alpha _j}(x_{S_j}) = \alpha _j(x_{S_j})\bigwedge _{x' \in \{x_{S_j}\}}\mu_{x' \rightarrow \alpha _j}(x'),
\end{align}

As stated in \cite{Frey, Aji}, after sufficient number of messages have been passed, the state of a variable node $x_i$ will be equal to $\beta_i(x_i)$. 

To obtain the correct value of the required marginal(s), it is essential that the factor graph be free of cycles. If there are cycles these may not be the correct values. The cycles can be eliminated by \textit{stretching} variable nodes or \textit{clustering} variable or factor nodes (refer to \cite[Sec. VI]{Frey} for a detailed description). These methods are exemplified in Section III-D. 

Both these graph transformations lead to enlargement of the local domain(s), and hence the configuration space of the node(s). In the new graph, the directed message passed from a vertex $v$ to $w$ is 
\begin{equation}
\label{eqn_msg_mod}
\mu _{v\rightarrow w}(x_{S_v\cap S_w})=\bigvee _{\mathclap{\{x_{S_v\backslash S_w}\}}}\;\alpha _v(x_{S_v})\;\bigwedge _{\mathclap{v'\in N(v)\backslash w}}\;\mu _{v'\rightarrow v} (x_{S_{v'}\cap S_v}),
\end{equation}
where $N(v)$ is the set of neighboring vertices of $v$ and its state $\sigma_v(x_{S_v})$ is
\begin{align} 
\label{eqn_state_mod}
\sigma_v(x_{S_v}) = \alpha _v(x_{S_v})\bigwedge _{v' \in N(v)}\mu_{v'\rightarrow v}(x_{S_{v'}\cap S_v})
\end{align}
These are the general forms of messages and states; \eqref{eqn_msg1}-\eqref{eqn_state_f} can be obtained from these.

Let $v^*$ be the node with the largest configuration space $A_{v^*}$ (choose any one if there are multiple such nodes). Then the number of operations required for computing messages and states in the SP algorithm will be $\mathcal{O}(A_{v^*})$. Thus, at the cost of possibly increased computational cost, the SP algorithm on the transformed graph yields the exact value of the marginals. In the sequel, we assume that the factor graph is acyclic.


\section{Decoding Network Codes using the Sum-Product Algorithm}
In this section, we show that decoding a NC is an MPF problem over a Boolean semiring. We provide a method to construct factor graph for decoding at a sink node using the SP algorithm. 

Though the factor graph approach and the junction tree approach are equivalent formulations to solve MPF problems, we prefer the former because of the amount of preprocessing required to obtain junction tree as argued below:
\begin{enumerate}
\item The construction of a junction tree for an MPF problem requires: \textit{(a)} construction of a \textit{moral graph}, \textit{(b)} its \textit{minimum complexity triangulation} if it is not already triangulated, \textit{(c)} construction of the \textit{clique graph} of the triangulated moral graph, and \textit{(d)} finding a spanning tree which leads to minimum computational cost. To the nodes of this clique tree the local kernels and variables of the MPF problem are attached \cite{Aji} to obtain the junction tree (a kernel or a variable is attached to a node of clique tree iff its local domain is a subset of the local domain of the said clique tree node). Thus, the GDL always gives the exact solution of the MPF problems.
\item A factor graphs is easily described by the local kernels associated with the MPF problem; it is a bipartite graph involving set of variable and a set of local kernels of the MPF kernels as the two vertex sets. If it is acyclic, then the SP algorithm gives the exact solution, if not, it gives an approximate solution. The SP algorithm is known to perform well even if the factor graph has cycles, for example, in decoding of LDPC and turbo codes. Factor graphs with cycles can be transformed into acyclic ones to obtain exact solutions \cite[Sec.~VI]{Frey}.
\end{enumerate}


\subsection{NC Decoding as an MPF Problem}
Given an acyclic network $N=(V,E)$, the demands at each sink, $D_k,\,k\in [K]$ and a set of global encoding maps, $\{\tilde{f}_e:e\in E\}$, that satisfy all the sink demands, the objectives at a sink, say $k^{th}$, is to find the instance of $x_{D_k}$ that was generated by the source(s) using the data it receives on its incoming edges, \emph{i.e.},
\begin{align}
\label{eqn_mpf}
x_{D_k}^* &= \underset{D_k}{\mathtt{supt}} \bigvee _{x_{[\omega]}\in F^{\omega}} \underbrace{\bigwedge _{e \in In(T_k)} \delta \left( \tilde{f}_e (x_{[\omega]})\, , \, y_e \right)}_{\beta ^{(k)}(x_{[\omega]})} 
\end{align}
Here $\beta ^{(k)}$ is the global kernel of the MPF problem at the $k^{th}$ sink and $\delta$ is a function that indicates whether its two input arguments are equal or not,\textit{i.e.} ,
\[
 \delta(a,b) =
  \begin{cases}
   0, & \text{if}\; a \neq b \\
   1, & \text{if}\; a = b
  \end{cases}
\] 
For LNC, \eqref{eqn_mpf} becomes
\begin{align} \nonumber
x_{D_k}^*& = \underset{D_k}{\mathtt{supt}} \bigvee _{\mathbf{x}\in F^\omega}\; {\bigwedge _{e\in In(T_k)}\delta \left( \mathbf{f}_e \cdot \mathbf{x}\, , \, y_e \right)}
\end{align}

Thus, decoding a NC has the form of a special class of MPF problems over Boolean semiring wherein we are interested only in some coordinates (specified by $D_k$) of the $\omega$-tuples in the support set and not the value of the global kernel. 

Since the solution $x_{D_k}^*$ is unique, individual coordinates $j\in D_k$ can be separately computed, \emph{i.e.}, 
\begin{equation}
\label{eqn_marg1}
\begin{aligned} 
x_j^* = \underset{j}{\mathtt{supt}} \bigvee _{x_j\in F}\;\beta _j^{(k)}(x_j) \\ 
\beta _j^{(k)}(x_j) = \bigvee _{\{x_{[\omega]\backslash j}\}}\beta ^{(k)}(x_{[\omega])},
\end{aligned}
\end{equation}
where $\beta _j^{(k)}(x_j)$ is the $j^{th}$ marginalization of the global kernel $\beta ^{(k)}$.

\begin{figure}[h]
\centering
\includegraphics[scale=0.5]{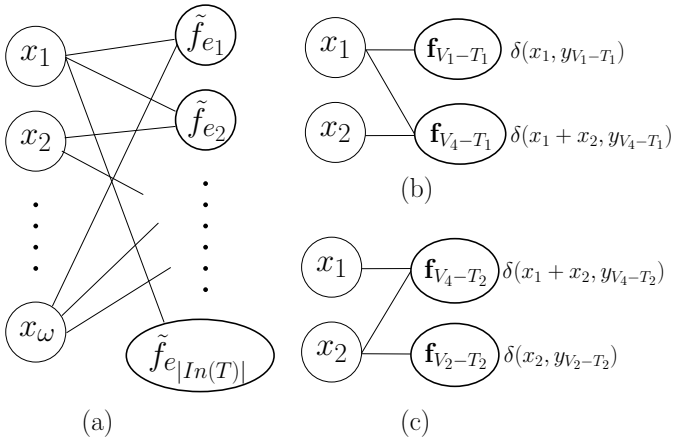}
\caption{(a) General form of a factor graph (b) Factor graphs for $T_1$ and (c) $T_2$ of the butterfly network. Local kernels are given adjacent to the function vertices.}
\label{junc_grph}
\vspace{-7pt}
\end{figure}

The factor graph for decoding at sink $T_k,k\in [K]$ is constructed as follows:
\begin{enumerate}
\item Install $\omega$ \textit{variable nodes}, one for each source message. These vertices are labeled by their corresponding source messages, $x_i$.
\item Install $|In(T_k)|$ \textit{factor nodes} and label them $\tilde{f}_e, e\in In(T_k)$. The associated local domain of each such vertex is the set, $S\subseteq x_{[\omega]}$, of source messages that participate in that encoding map and the local kernel is $\delta(\tilde{f}_e(x_{[\omega]})\, , \, y_e)$. These vertices are labeled by their corresponding local kernels, $\tilde{f}_e$.
\item A variable node is connected to a factor node iff the source message corresponding to that variable node participates in the encoding map corresponding to the said factor node.
\end{enumerate}
We use thicker lines for factor nodes to differentiate them from variable nodes. The factor graph so constructed will be a bipartite graph. General form of a factor graph and the same for the two sink nodes of the butterfly network are given in Fig.~\ref{junc_grph} (cf. \cite[Fig. 3]{Salmond}).

Messages and states are computed using \eqref{eqn_msg_mod} and \eqref{eqn_state_mod} respectively. As stated before, once a node (say $v$), has received message from all the adjacent nodes, its state, $\sigma _v(x_{S_v})$, can be computed. Let $x_{D_k}\cap x_{S_v}= x_{B}$. The value of the subset $B$ of the requisite source messages at the $k^{th}$ sink node is
\begin{equation}
x_B^*=\underset{B}{\mathtt{supt}} \bigvee _{\{x_{S_v}\}} \alpha _v(x_{S_v})\bigwedge _{v'\in N(v)} \mu _{v'\rightarrow v} (x_{S_{v'}\cap S_v})
\end{equation}

As specified in Section II, the SP algorithm yields the correct value of the source messages if the factor graph is a tree. If not then the cycles in the factor graph will have to be removed.

\subsection{Traceback}
Since we are interested in the value of the source messages and not the value of the marginalizations of the global kernel, we can use \emph{traceback} \cite{LP} to further lower the number of computations.

Assume that the single-vertex SP algorithm is run with vertex $v$ as the root (all messages are directed towards $v$) and the value $x_{S_v}^*\subseteq x_{D_k}$ of some source messages in the demand set of the $k^{th}$ sink has been ascertained. Now, partition the local domain of a neighboring node $w$, as $x_{S_w}=x_A \cup x_B$, where $A=S_w\backslash S_v$ and $B=S_w\cap S_v$. Since $x_{S_v}^*$ is known, the value $x_B^*$ that causes $\sigma_w(x_A,x_B)$ to take value $1$ is also known. The value of the source messages $x^*_A$ can then be obtained using \eqref{eqn_state_mod} as follows:
\begin{equation}
\begin{aligned} 
\label{eqn_trcbk}
x^*_A & =\underset{A}{\mathtt{supt}} \bigvee _{\{x_A\}} \sigma_w(x_A,x_B^*)\\
& = \underset{A}{\mathtt{supt}} \bigvee _{\{x_A\}} \mu_{v\rightarrow w}(x_B^*)\; \lambda_w(x_A,x_B^*)\\
 & = \underset{A}{\mathtt{supt}} \bigvee_{\{x_A\}} \lambda_w(x_A,x_B^*),
\end{aligned}
\end{equation}
where
\[\lambda_w(x_A,x_B)=\alpha (x_{S_w})\bigwedge_{v'\in N(w)\backslash v} \mu_{v'\rightarrow w}(x_{S_{v'}\cap S_w})\]
is the partial state computed at $w$ while passing the message $\mu_{w \rightarrow v}(x_B)$ to the root $v$. Thus, $\mu_{v\rightarrow w}(x_B^*)$ does not need to be computed or passed, leading to saving of operations. 

Here we have exploited the fact that we require only $\mathtt{supt}_A \bigvee _{x_A} \sigma_w$ and not the value of $\sigma_w$ which would have required passing of the message $\mu_{v\rightarrow w}(x_B^*)$ from $v$ to $w$ too. Hence, the traceback step reduces the computational complexity.

The traceback step can be used repeatedly until values of all the source messages in $x_{D_k}$ are obtained. This can be done by using \eqref{eqn_trcbk} on other neighbors of $v$ and then neighbors of neighbors of $v$ and so on. This can lead to considerable reduction in number of operations and is exemplified in Section III-D.

\begin{figure*}[t]
\centering
\includegraphics[scale=0.5]{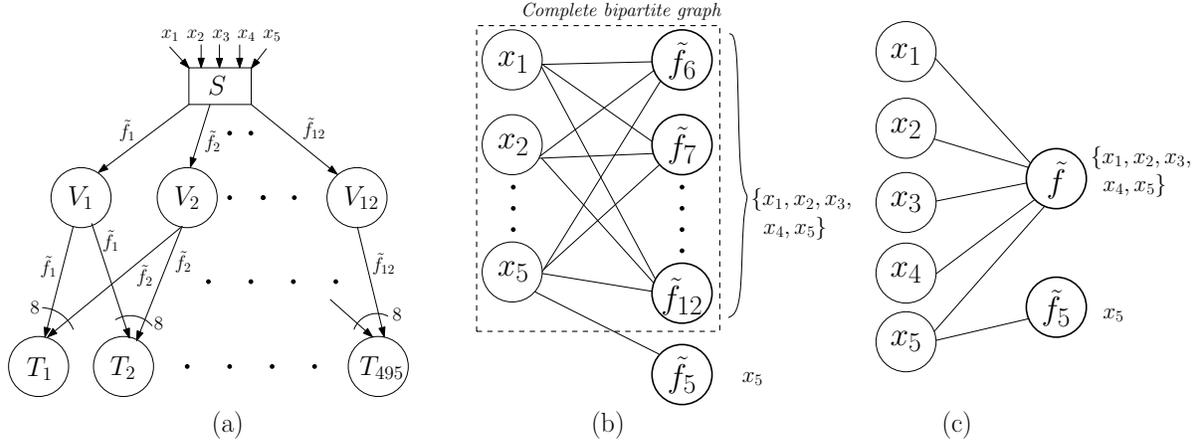}
\caption{(a) Combination network (b) Factor graph with cycles for $T_{495}$ and (c) one obtained by clustering equation nodes $\tilde{f}_6-\tilde{f}_{12}$ $(\tilde{f}=\bigwedge_{i=6}^{12}\tilde{f}_i)$}
\label{trcbk}
\end{figure*}

\subsection{Computational Complexity}
We suggest using SP algorithm for decoding a network code only when the code is either non-linear or it is linear but the number of messages is very large. For linear network codes with manageable value of $\omega$, Gaussian elimination with backward substitution is advisable. If using SP algorithm for decoding network codes (when warranted) leads to computational complexity strictly better than brute-force decoding complexity, then the code is called \emph{fast decodable network code}.

We now discuss the computational complexity of SP algorithm based decoding of network codes. As in \cite{Frey}, by complexity we mean the number of semiring operations required to obtain the desired source message. For convenience, we assume that all the source messages take value from a $q$-ary alphabet; results for variable alphabet size can be obtained similarly. 

As stated above, in order to recover the requisite source messages at a sink we need only run single-vertex SP algorithm followed by traceback steps. For a given sink node, if the factor graph constructed using the method given in Section III-A is cycle-free and the network code is such that the local domains of all factor nodes have cardinality at most $l\, (<\omega)$, then the number of operations required for decoding using the SP algorithm is $\mathcal{O}(q^l)$. If the sink demands all the source messages, then the brute-force decoding would require $\mathcal{O}(q^{\omega}) (> \mathcal{O}(q^l))$ operations. Thus, an acyclic factor graph with at most $l\, (<\omega)$ variables per equation is a sufficient condition for fast decodability of the network code at a sink which demands all the source messages.

If the graph is not cycle-free then we remove the cycles using the methods specified above and let $m\leqslant \omega$ be the size of maximum cardinality local domain in the transformed cycle-free factor graph. The number of computations required now will be $\mathcal{O}(q^m)\leqslant \mathcal{O}(q^{\omega})$ and the code is fast decodable iff $m<\omega$.

\subsection{Illustrations}
We now present some examples illustrating use of the SP algorithm to decode NC.

\begin{example}
Consider the butterfly network of Fig.\ref{b_fly}. Here $q=\omega =2$. The factor graphs for two sink nodes are given in are given in Fig.~\ref{junc_grph}(b) and (c). The messages passed and state computations for decoding at $T_1$ are as follows:
\begin{align} \nonumber
\mu_{x_2\rightarrow \mathbf{f}_{V_4-T_1}}(x_2)&= 1 \\ \nonumber
\mu_{\mathbf{f}_{V_1-T_1}\rightarrow x_1}(x_1)&= \delta(x_1,y_{V_1-T_1}) \\   \nonumber
\mu_{\mathbf{f}_{V_4-T_1}\rightarrow x_1}(x_1)&= \bigvee _{x_2}\delta(x_1+x_2,y_{V_4-T_1})  \\
\nonumber
\mu_{x_1\rightarrow \mathbf{f}_{V_4-T_1}}(x_1)&= \mu_{\mathbf{f}_{V_1-T_1},x_1}(x_1)  \\ \nonumber
\sigma_{x_1}(x_1)&= \mu_{\mathbf{f}_{V_1-T_1},x_1}(x_1) \; \mu_{\mathbf{f}_{V_4-T_1}\rightarrow x_1}(x_1) 
\end{align}
\begin{align}
 \nonumber
\mu_{\mathbf{f}_{V_4-T_1}\rightarrow x_2}(x_2)&= \bigvee _{x_1}\delta(x_1+x_2,y_{V_4-T_1})\; \mu_{\mathbf{f}_{V_1-T_1}\rightarrow x_1}(x_1)  \\ \nonumber
\sigma_{x_2}(x_2)&= \mu_{\mathbf{f}_{V_4-T_1},x_2}(x_2)
 \end{align}
It is easy to verify that $\sigma_{x_i}(x_i)=\beta_i(x_i), i=1,2$. Hence, $x_i^*=\mathtt{supt}_{x_i} \bigvee _{x_i}\sigma_{x_i}(x_i), i=1,2$.
Similar computations apply for $T_2$ also. This network code is not fast decodable since the number of computations required is $\mathcal{O}(q^{\omega})$. \hfill $\square$
\end{example}

Now we present a multicast network which admits no linear solution over $\mathbb{F}_2$, but a non-linear solution exists.

\begin{example}
Consider the multicast combination network given in Fig.\ref{trcbk}(a). All messages take value from $F_2$. Each sink is connected to a distinct size $8$ subset of intermediate node. Multicast, in $F_2$, is feasible over such a network iff a $(12,32,5)$ binary error correcting code exists \cite{Riis}. The Nadler's code is one such systematic code with the requisite parameters. Note that since there exist no $(12,32,5)$ binary linear code, the above multicast network, admits no linear solution over $F_2$.
\begin{figure*}[t]
\centering
\includegraphics[scale=0.5]{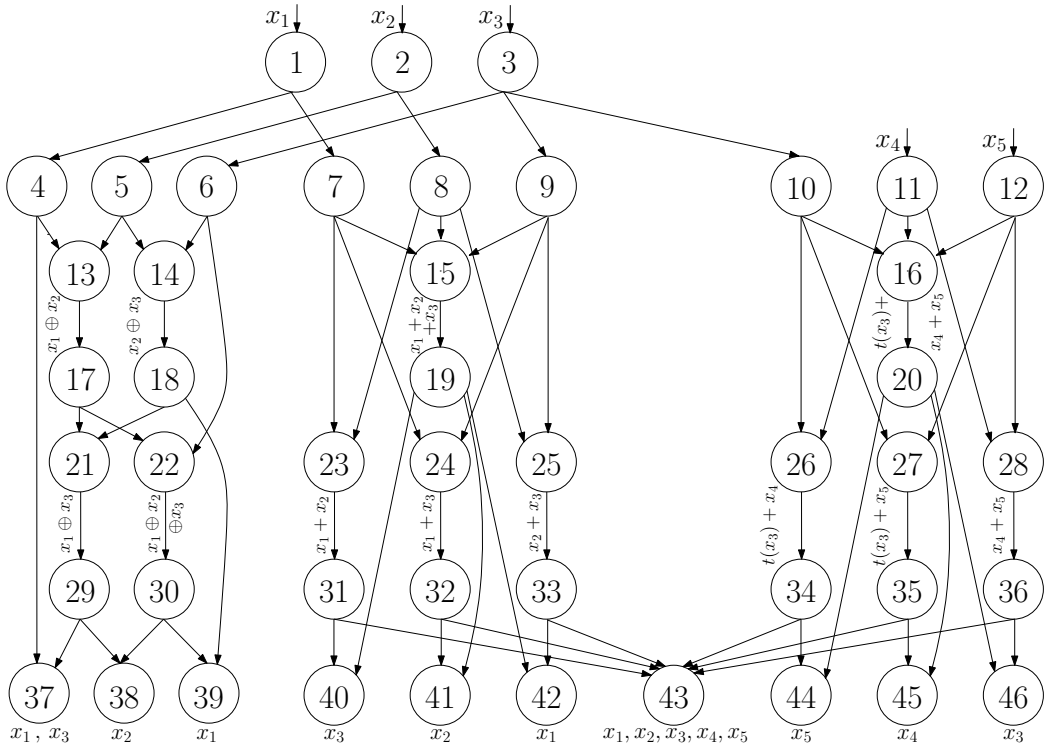}
\caption{The network $\mathcal{N}_3$ of \cite{Insuff}.}
\label{fano}
\end{figure*}

Apart from the systematic part of the Nadler's code, the $7$ redundant bits are encoded using non-linear functions \cite{Wet};these are:
\begin{align}
\nonumber \tilde{f}_i &= x_i, \;\;\; i\in [5] \\
\nonumber \tilde{f}_6 &= x_1+x_2+x_3+(x_1+x_5)(x_3+x_4) 
\\
\nonumber \tilde{f}_7 &= x_1+x_2+x_4+(x_1+x_3)(x_4+x_5) 
\\
\nonumber \tilde{f}_8 &= x_1+x_2+x_5+(x_1+x_4)(x_3+x_5) 
\\
\nonumber \tilde{f}_9 &= x_2+x_3+x_4+x_1x_4+x_4x_5+x_5x_1 \\
\nonumber \tilde{f}_{10} &= x_2+x_3+x_5+x_1x_3+x_3x_5+x_5x_1 
\\
\nonumber \tilde{f}_{11} &= x_2+x_4+x_5+x_1x_3+x_3x_5+x_5x_1 \\
\nonumber \tilde{f}_{12} &= x_1+x_2+x_3+x_4+x_5+ x_3x_4+ x_4x_5+ x_5x_3 
\end{align}
These functions are the global encoding maps of the $12$ source to intermediate node links. The intermediate nodes simply route the data on incoming edges to the connected sink nodes, hence, the global encoding maps of the incoming and outgoing edges of an intermediate node is same. 

Consider decoding at the $495^{th}$ sink whose global encoding maps of incoming edges are $\tilde{f}_i,\;i=5,\ldots,12$. The factor graph constructed using the method stipulated above will have cycles (Fig.~\ref{trcbk}(c)). To eliminate the cycles, we cluster $\tilde{f}_i,\;i=6,\ldots,12$ function vertices into a single one with the local kernel as the product of $7$ original ones (Fig.~\ref{trcbk}(c)). The computational complexity of SP decoding is same as that of brute-force decoding ($\mathcal{O}(q^5)$). This is true for all the sink nodes and hence this code is not fast decodable for any of the sinks. \hfill $\square$
\end{example}

In the next example we present a network with general demands at sinks and employ the SP algorithm for decoding a vector non-linear network code for it. We also demonstrate usefulness of traceback for saving computations of some messages in the factor graph.

\begin{example}
Consider the network given in Fig.~\ref{fano}. The sinks (nodes $37-46$) have general demands which are specified by variables below them. In \cite{Insuff}, the authors showed that this network admits no linear solution over any field and gave a vector non-linear solution. The source message $x_i,i\in[5]$ are 2-bit binary words ($q=4,\,\omega = 5$), $+$ denotes addition in ring $\mathbf{Z}_4$, $\oplus$ denotes the bitwise XOR and the function $t(\cdot)$ reverses the order of the 2-bit input. 

The factor graphs for nodes $37$, $40$ and $43$, denoted by $\mathcal{G}_{37}, \mathcal{G}_{40}$ and $\mathcal{G}_{43}$ respectively, are given in Fig.~\ref{fano_junc}; the same for other nodes have similar structure and can be constructed using method given in Section III-A. Note that $\mathcal{G}_{40}$ has a cycle of length $4$ and $\mathcal{G}_{43}$ has two cycles of length $6$ each. The cycles are removed as follows:
\begin{itemize}
\item The cycle in $\mathcal{G}_{40}$ is removed by clustering the variable vertices $x_1$ and $x_2$; the local domain and kernel of the new variable vertex are $(x_1,x_2)$ (union of the local domains of the clustered vertices) and  $1$ (product of the local kernels of the clustered vertices) (Fig.~\ref{fano_junc}).
\item The cycles $C_1$ and $C_2$ in $\mathcal{G}_{43}$ are removed by deleting the dotted edges and \textit{stretching} variable vertex $x_3$ around the respective cycles; this involves adding the stretched variable to all the local domain in the cycle and leaving the local kernels unchanged (Fig.~\ref{fano_junc}).
\end{itemize}

\begin{figure*}[t]
\centering
\includegraphics[scale=0.5]{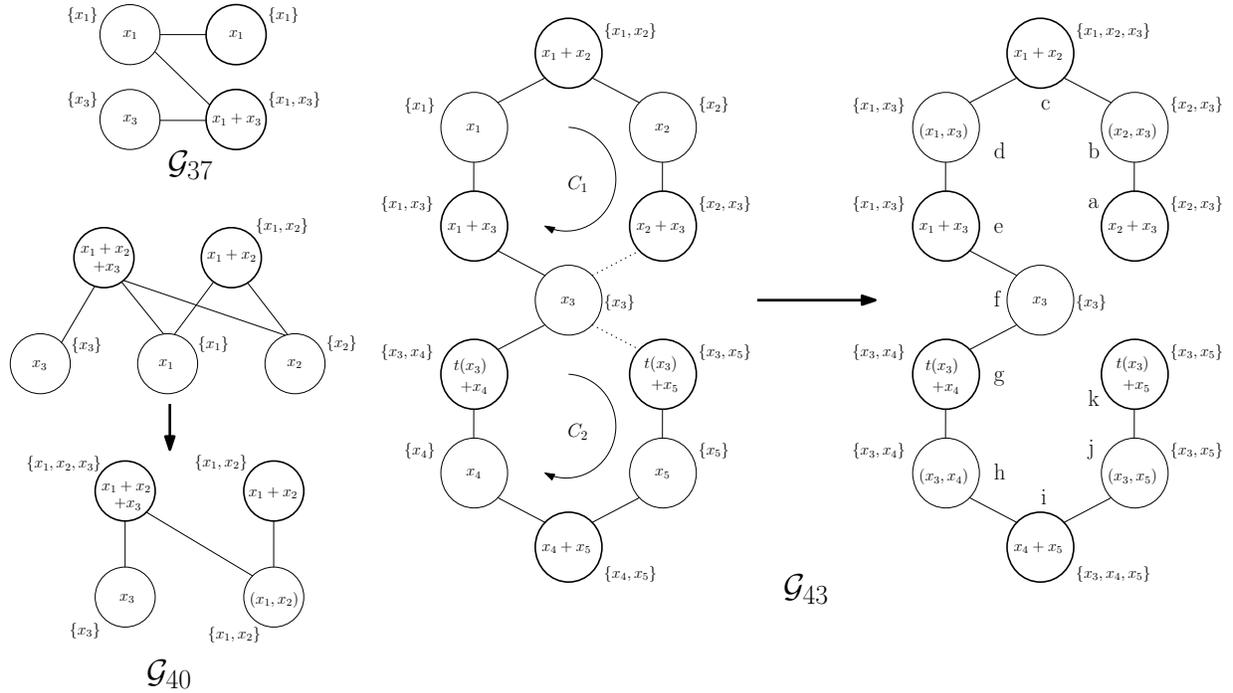}
\caption{The factor graphs for sinks with labels $37$, $40$ and $43$ in network of Fig.~\ref{fano}.}
\label{fano_junc}
\end{figure*}
We infer from $\mathcal{G}_{43}$ that the number of computations required to reproduce all the source messages at $V_{43}$ is only $\mathcal{O}(q^3)$ instead of $\mathcal{O}(q^5)$ (as brute-force decoding would have required) and hence this code is fast decodable for $V_{43}$. The decoding process at $V_{43}$, using single-vertex SP algorithm with node ``f'' in $\mathcal{G}_{43}$ as root followed by traceback to compute $x_1,x_2,x_4,x_5$, is demonstrated below:
\begin{align}
\nonumber & \mu _{k\rightarrow j}=\mu _{j\rightarrow i}(x_3,x_5)=\delta (t(x_3)+x_5,y_{35-43}) 
\\
\nonumber & \mu _{a\rightarrow b}=\mu _{b\rightarrow c}(x_2,x_3)=\delta (x_2+x_3,y_{33-43}) 
\\
\nonumber & \mu _{i\rightarrow h}=\mu _{h\rightarrow g}(x_3,x_4)= \bigvee _{x_5} \delta (x_4+x_5,y_{36-43}) \mu _{j\rightarrow i}(x_3,x_5) 
\\
\nonumber & \mu _{c\rightarrow d}=\mu _{d\rightarrow e}(x_1,x_3)= \bigvee _{x_2} \delta (x_1+x_2,y_{31-43}) \mu _{b\rightarrow c}(x_1,x_2)
\\
\nonumber & \mu _{e\rightarrow f}(x_3)=\bigvee _{x_1}\delta (x_1+x_3,y_{32-43})\mu _{d\rightarrow e}(x_1,x_3)
\\
\nonumber & \mu _{g\rightarrow f}(x_3)=\bigvee _{x_4}\delta (t(x_3)+x_4,y_{34-43})\mu _{h\rightarrow g}(x_3,x_4)
\end{align}
At ``f'', decoding of $x_3$ is performed as follows:
\begin{align}
\nonumber x_3^*&=\underset{x_3}{\mathtt{supt}}\bigvee _{x_3}\sigma _f(x_3) =\underset{x_3}{\mathtt{supt}}\,\sigma _f(x_3) \\\nonumber &= \underset{x_3}{\mathtt{supt}}\,\mu _{e\rightarrow f}(x_3)\,\mu _{g\rightarrow f}(x_3)
\end{align}
We can now use traceback \eqref{eqn_trcbk} to compute $x_1$ and $x_2$ without having to compute $\mu _{f\rightarrow e},\mu _{e\rightarrow d}$ or $\mu _{d\rightarrow c}$ as follows:
\begin{align}\nonumber
x_1^*&=\underset{x_1}{\mathtt{supt}}\bigvee _{x_1}\sigma _e(x_1,x_3^*) = \underset{x_1}{\mathtt{supt}}\,\lambda _e(x_1,x_3^*)\\ \nonumber & =\underset{x_1}{\mathtt{supt}}\, \delta (x_1+x_3^*,y_{32-43})\mu _{d\rightarrow e}(x_1,x_3^*)
\\
\nonumber
x_2^*&=\underset{x_2}{\mathtt{supt}}\bigvee _{x_2}\sigma _c(x_1^*,x_2,x_3^*) = \underset{x_2}{\mathtt{supt}}\,\lambda _c(x_1^*,x_2,x_3^*)\\ \nonumber & =\underset{x_2}{\mathtt{supt}}\, \delta (x_1^*+x_2,y_{31-43})\mu _{b\rightarrow c}(x_2,x_3^*)
\end{align}
Note that $\mu _{d\rightarrow e}$ and $\mu _{b\rightarrow c}$ were already computed. Without traceback, $x_1$ and $x_2$ are decoded at variable nodes ``d'' and ``b'' respectively as follows: 
\begin{align*}
\mu _{f\rightarrow e}(x_3)&=\mu _{g\rightarrow f}(x_3)\\
\mu _{e\rightarrow d}(x_1,x_3)&=\mu _{f\rightarrow e}(x_3) \delta (x_1+x_3,y_{32-43})\\
x_1^*&=\underset{x_1}{\mathtt{supt}}\bigvee _{x_3}\,\mu _{e\rightarrow d}(x_1,x_3) \mu _{c\rightarrow d}(x_1,x_3)\\
\mu _{d\rightarrow c}(x_1,x_3)&=\mu _{e\rightarrow d}(x_1,x_3)\\
\mu _{c\rightarrow b}(x_2,x_3)&= \bigvee _{x_1}\,\mu _{d\rightarrow c}(x_1,x_3)\delta (x_1+x_2,y_{31-43})\\
x_2^*&=\underset{x_2}{\mathtt{supt}}\bigvee _{x_3}\,\mu _{c\rightarrow b}(x_2,x_3) \mu _{a\rightarrow b}(x_2,x_3)
\end{align*}
Similarly $x_4$ and $x_5$ can be obtained. The number of semiring operations required to compute all the messages passed and states computed is given in Table~I.

\begin{table}[ht] 
\normalsize
  \centering

  \begin{tabular}{>{\centering}m{0.2in} >{\centering}m{1in} >{\centering}m{0.6in} >{\centering\arraybackslash}m{0.6in}}
    \toprule
  & \textbf{Messages/States} & \textbf{No. of $\bigwedge$} & \textbf{No. of $\bigvee$} \\
    \midrule
C1 & $\mu _{k\rightarrow j},\mu _{j\rightarrow i}$ & $0$ & $0$ \\
C2 & $\mu _{a\rightarrow b},\mu _{b\rightarrow c}$ & $0$ & $0$ \\
C3 & $\mu _{i\rightarrow h},\mu _{c\rightarrow d}$ & $q^3$ & $q^2(q-1)$ \\
C4 & $\mu _{h\rightarrow g},\mu _{d\rightarrow e}$ & $0$ & $0$ \\
C5 & $\mu _{e\rightarrow f},\mu _{g\rightarrow f}$ & $q^2$ & $q(q-1)$ \\ \hline
C6 & $x_3^*$ & $q$ & $0$ \\
C7 & $x_1^*,x_4^*$ & $q$ & $0$ \\
C8 & $x_2^*,x_5^*$ & $q$ & $0$ \\ \hline
C9 & $\mu _{f\rightarrow e},\mu _{f\rightarrow g}$ & $0$ & $0$ \\
C10& $\mu _{e\rightarrow d},\mu _{g\rightarrow h}$ & $q^2$ & $0$ \\
C11& $x_1^*$ & $q^2$ & $q(q-1)$ \\
C12& $\mu _{d\rightarrow c},\mu _{h\rightarrow i}$ & $0$ & $0$ \\
C13& $\mu _{c\rightarrow b},\mu _{i\rightarrow j}$ & $0$ & $0$ \\
C14& $x_2^*$ & $q^2$ & $q(q-1)$ \\
    \bottomrule
    \end{tabular}
    \vspace{5pt}
\caption{}  
\vspace{-20pt}
  \end{table}

The number of computations required with traceback are 2 (C1+C2+$\ldots$+C5)+C6+C7+C8 which is $2(q^3+q^2)+5q=180$ products ($\wedge$) and $2(q^2(q-1)+q(q-1))=120$ sums ($\vee$).Without traceback, the number of operations required are 2(C1$\ldots$+C5)+C6+2(C9+C10)+C11+2(C12+C13)+C14 which is $2q^3+6q^2=224$ products  and $2q^2(q-1)+4q(q-1)=144$ sums. Thus, running single-vertex SP algorithm followed by traceback step affords computational advantage over multiple-vertex version.

\hfill $\square$
\end{example}

\section{Discussion}
In this paper, we proposed a SP algorithm based decoder for decoding NC. Subsequently, a method for constructing the factor graph for a given sink node using the global encoding maps (or vectors in case of LNC) of the incoming edges and demands of the sink was provided. The graph so constructed had fewer nodes and led to fewer message being passed lowering the number of operations as compared to the scheme of \cite{Salmond}. Next we discussed how cycles in factor graph affect the solution of the MPF problem and illustrated with examples how to circumvent them. We introduced and discussed the advantages of traceback over multiple-vertex SP algorithm. Next, for the sinks demanding all the source messages, we introduced the concept of fast decodable network codes and provided a sufficient condition for a network code to be fast decodable.



\end{document}